\documentclass[journal=nalefd, layout=traditional, manuscript=letter]{achemso}

\usepackage{graphicx}
\usepackage{dcolumn}
\usepackage{bm}
\usepackage{caption}
\usepackage{subcaption}
\usepackage{amssymb,amsmath}
\usepackage{gensymb}
\usepackage{url}
\usepackage{float}
\usepackage{hyperref}    
\usepackage{cleveref}[2012/02/15]

\let\oldmaketitle\maketitle
\let\maketitle\relax

\author{Rahul Debnath}
\altaffiliation{Contributed equally to this work}
\email{rahuldebnath@iisc.ac.in}
\author{Indrajit Maity}
\altaffiliation{Contributed equally to this work}
\author{ Manish Jain}
\affiliation{Department of Physics, Indian Institute of Science, Bangalore 560012, India}
\author {Arindam Ghosh}
\vspace{1.5cm}
\affiliation{Department of Physics, Indian Institute of Science, Bangalore 560012, India}
\alsoaffiliation{Centre for Nano Science and Engineering, Indian Institute of Science, Bangalore 560012, India}
\title {Evolution of high-frequency Raman modes and their doping dependence in twisted bilayer $\mathrm{MoS_{2}}$}


\begin{document}
	\oldmaketitle
\begin{abstract}
	Twisted van der Waals heterostructures unravel a new platform to study strongly correlated quantum phases. The interlayer coupling in these heterostructures is sensitive to twist angles ($\theta$) and key to controllably tune several exotic properties. Here, we demonstrate a systematic evolution of the interlayer coupling strength with twist angle in bilayer $\mathrm{MoS_{2}}$ using a combination of Raman spectroscopy and classical simulations. At zero doping, we show a \textit{monotonic} increment of the separation between the $\mathrm{A_{1g}}$ and $\mathrm{E^{1}_{2g}}$ mode frequencies as $\theta$ decreases from $10^{\circ} \to 1^{\circ}$, which saturates to that for a bilayer at small twist angles. Furthermore, using doping-dependent Raman spectroscopy we reveal $\theta$ dependent softening and broadening of the $\mathrm{A_{1g}}$ mode, whereas the $\mathrm{E^{1}_{2g}}$ mode remains unaffected. Using first principles based simulations we demonstrate large (weak) electron-phonon coupling for the $\mathrm{A_{1g}}$ ($\mathrm{E^{1}_{2g}}$) mode explaining the experimentally observed trends. Our study provides a non-destructive way to characterize the twist angle, the interlayer coupling and establishes the manipulation of phonons in twisted bilayer $\mathrm{MoS_{2}}$ (twistnonics).

\end{abstract}
	
\section{Introduction}
 The choice of the materials, number of layers and relative rotation between the layers (\textit{twist}) provide three important degrees of freedom to engineer the properties of the van der Waals (vdW) heterostructures \cite{pisoni2017gate,wang2012electronics,cui2015multi,sarkar2015subthermionic,roy2013graphene,ahmed2019thermodynamically,naik2017origin}. The introduction of a twist between two layers of two dimensional materials generates a large-scale periodic pattern, called as moir\'{e} superlattice. In the case of twisted bilayer graphene the moir\'{e} superlattice induced periodic potential for electrons yield many fascinating phenomena such as, flat bands in the electronic band structure which can host correlated insulating phase \cite{cao2018correlated}, superconductivity \cite{cao2018unconventional,yankowitz2019tuning}, and ferromagnetism \cite{sharpe2019emergent}. The ability to tune these properties in a controlled manner requires detailed understanding of the evolution of interlayer coupling strength with twisting \cite{yankowitz2019tuning,yao2018quasicrystalline,kim2012raman,havener2012angle}.

\begin{figure*}[t]
\vspace*{-0.5cm}
\includegraphics[width=1\linewidth]{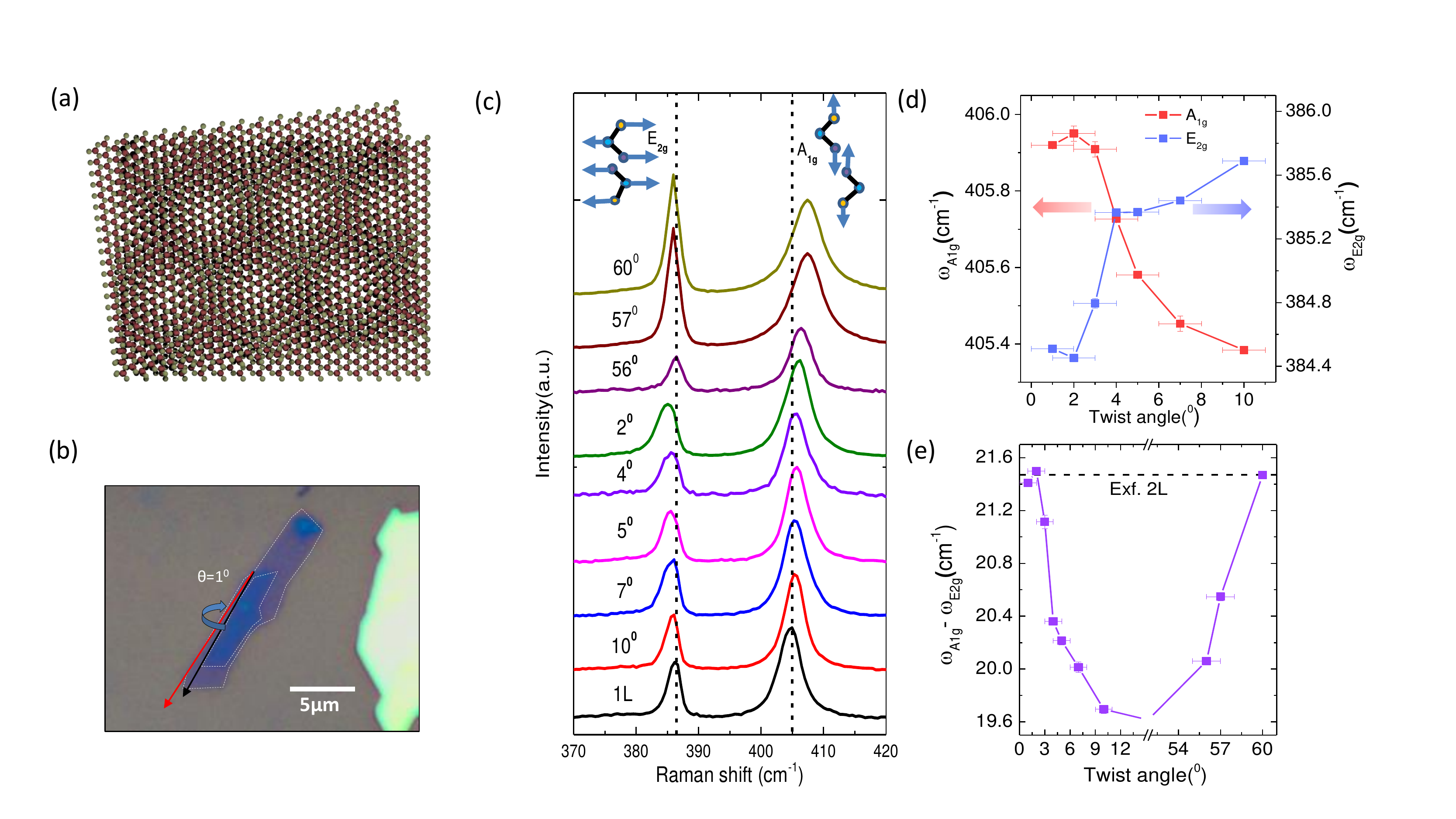}
\caption{(a) Schematic of a generic  moir\'e pattern. (b) Optical micropgraph of twisted bilayer MoS$_2$ with the twist angle defined as the relative rotation between two straight edges. (c) Raman spectra of monolayer MoS$_2$ and bilayer MoS$_2$ with the dashed lines representing phonon modes of monolayer MoS$_2$. (d),(e) Twist angle dependence of the individual $\mathrm{E^{1}_{2g}}$, $\mathrm{A_{1g}}$ mode frequencies and their peak separation, respectively. A dashed black line is shown for the peak separation of naturally exfoliated  bilayer MoS$_2$. The error bars show uncertainty in $\theta$ and peak position. }
\end{figure*}

Compared to graphene, the effects of twisting in $\mathrm{MoS_{2}}$ layers, an important material for electronic and optoelectronic applications \cite{roy2013graphene,sarkar2015subthermionic,wang2012electronics}, has been relatively less explored \cite{huang2014probing, lin2018moiré,liu2014evolution,huang2016low, puretzky2016twisted,zheng2015coupling,zhang2019tunable,van2014tailoring,naik2018ultraflatbands,zhu2018moire,yeh2016direct,Maity_ultrasoft_2019,wu2019topological,fleischmann2019moir,naik2019origin,lu2017twisted,tan2016first,zhou2017interlayer}. The existing experimental studies have predominantly concentrated on the change in optical properties and low frequency vibrational modes using chemical vapor deposition (CVD) grown samples \cite{huang2014probing, lin2018moiré,liu2014evolution,huang2016low, puretzky2016twisted,zheng2015coupling,zhang2019tunable,van2014tailoring}. Although, the change in the interlayer coupling strength with twist angle is evident in these studies, a definitive experiment relating the \textit{precise} evolution of the interlayer coupling strength with twist angle is still lacking. 

Electron-phonon coupling (EPC) affects many important properties in solids, including those in two dimensional atomic layer, for instance carrier mobility, and thermalization of hot carriers \cite{radisavljevic2013mobility}. Doping dependent Raman spectroscopy is often used to probe EPC in two dimensional materials \cite{das2008monitoring,pisana2007breakdown,chakraborty2012symmetry, Ponomarev_multivalley_2019, lu2017gate,parkin2016raman}. The evolution of the Raman active modes upon electron doping in twisted bilayer of transition metal dichalcogenide (TMDC) can shed light on the twist angle dependence of EPC. As the electron mobility in 2D TMDC devices is often limited by electron-phonon interaction \cite{radisavljevic2013mobility}, by tuning the EPC through twist angle one can, in principle, engineer the mobility in these systems.

\begin{figure*}[t]
\vspace*{-0.5cm}
\includegraphics[width=1\linewidth]{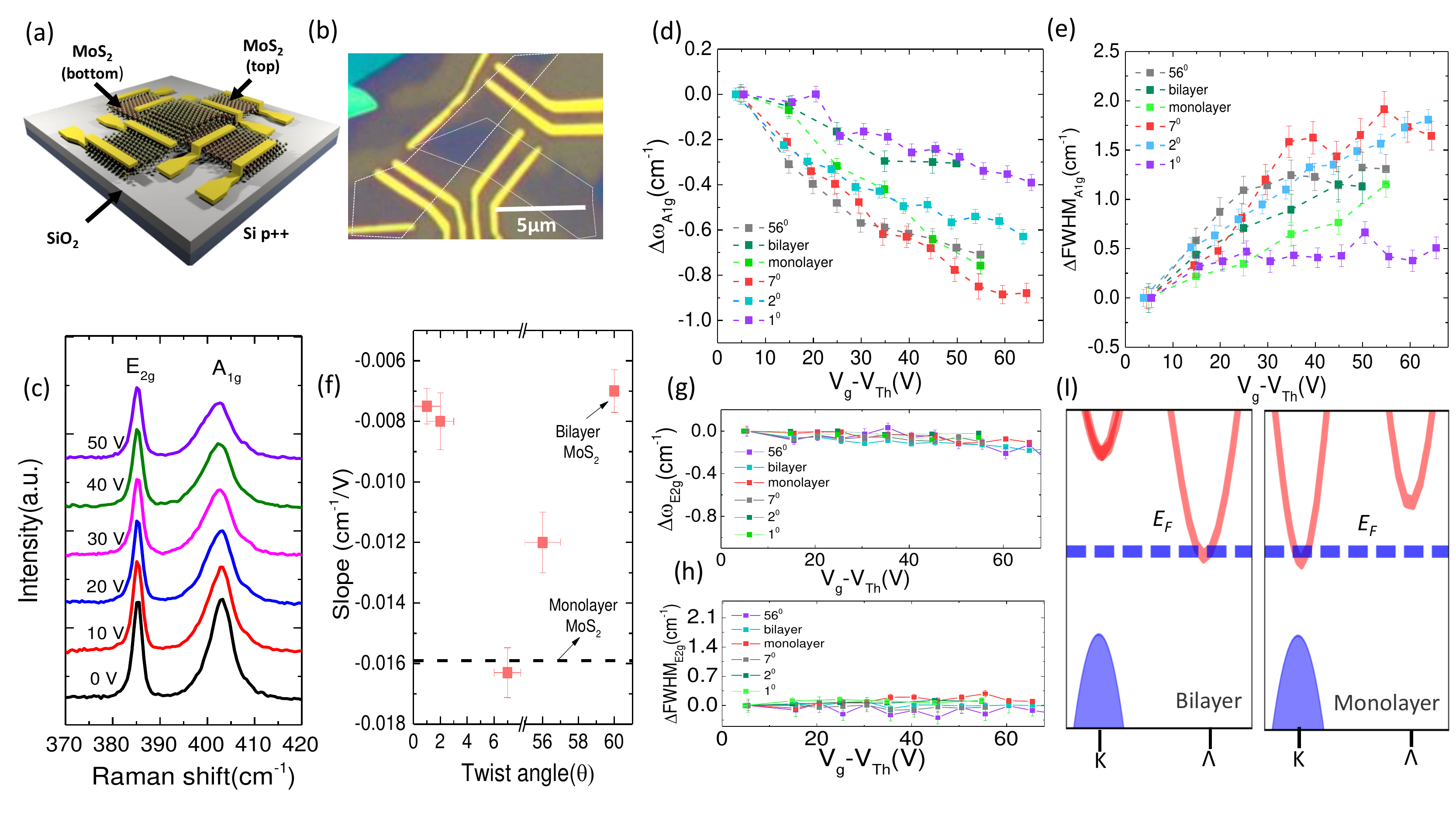}
\caption{(a) Device schematic of $\mathrm{tBLMoS_{2}}$ FET. (b) Optical micrograph of $\mathrm{tBLMoS_{2}}$ (${\theta}={56^\circ}$).(c) Raman spectra of $\mathrm{tBLMoS_{2}}$ for $\theta=7^\circ$ at different gate voltages. (d),(g) The gate voltage dependence of the softening of the $\mathrm{A_{1g}}$ and $\mathrm{E^{1}_{2g}}$ mode frequencies, respectively. (e),(h) The gate voltage dependence of the broadening (linewidth) of the $\mathrm{A_{1g}}$ and $\mathrm{E^{1}_{2g}}$ modes, respectively. (f) The slope for the softening of the $\mathrm{A_{1g}}$ mode frequency calculated by fitting a straight line to doping dependent Raman spectra. (i) Schematic of Fermi level position showing occupation of different valley in monolayer and bilayer $\mathrm{MoS_{2}}$ for maximum doping considered.}
\end{figure*}

In this letter, using bilayer MoS$_2$, a prototypical TMDC, we investigate the evolution of high-frequency phonon modes as a function of twist angle and electron doping using Raman spectroscopy. The twisted structures are prepared from mechanically exfoliated $\mathrm{MoS_{2}}$ layers, which exhibits higher mobility and lower disorder than CVD grown samples \cite{li2014preparation}. We demonstrate the \textit{monotonic} change of $\mathrm{E^{1}_{2g}}$ and $\mathrm{A_{1g}}$ modes for relatively small twist angles ($\theta \lesssim 10^{\circ}$) by combining Raman spectroscopy, classical and first principles based simulations. We also find two intriguing features of the electron doping dependence of these high-frequency modes. First, irrespective of number of layers (single or bilayer) or twist angle between them, the $\mathrm{A_{1g}}$ mode shows strong doping dependence; the phonon frequency decreases by $\sim$ 0.9 cm $^{-1}$ and the linewidth increases by $\sim 2\ \mathrm{cm^{-1}}$ for electron doping of $4.87 \times 10^{12}$/cm$^2$ at large twist angle. On the other hand, the phonon frequency and linewidth of the $\mathrm{E^{1}_{2g}}$ mode remain unchanged on electron doping. Second, the doping dependence of the $\mathrm{A_{1g}}$ mode frequencies for large (small) twist angles resemble to that of single layer (bilayer), thereby providing a new route to identify twist angle and doping concentration in twisted bilayer $\mathrm{MoS_{2}}$ ($\mathrm{tBLMoS_{2}}$). We explain these observations using first principles based calculations.

\section{Results and discussion}

\textit{Experiment:} Atomically thin crystalline layers of MoS$_2$ were prepared by standard mechanical exfoliation of  MoS$_2$ on 285nm SiO$_2$ on p$^{++}$-doped Si substrate using the scotch tape technique~\cite{novoselov2005two,novoselov2004electric}. We used dry transfer~\cite{roy2013graphene} method to stack two mechanically exfoliated MoS$_2$ to prepare twisted bilayer samples ($\mathrm{tBLMoS_{2}}$). By using micro-manipulation stage, this method allows the preparation of samples of the desired twist angle within the accuracy of 1$^{\circ}$. Additionally, this technique greatly reduces any contamination between the two layers, which ensures a robust and reproducible device properties. Fig.~1a,~1b show a generic moir{\'e} pattern due to twisting and an illustrative micrograph of $\mathrm{tBLMoS_{2}}$ after transfer on SiO$_2$/Si$^{++}$ substrate, respectively. Edge profiles of the top and bottom MoS$_2$ were determined using the technique described in Ref.~\cite{guo2016distinctive}. Among the 2D materials, MoS$_2$ shows a distinctive tendency to cleave along the zigzag direction. For each exfoliated monolayer flakes, angles between two straight edges (edge angle) were measured, and those with edge angle of $\approx 60^{\circ}$ were chosen. To make $\mathrm{tBLMoS_{2}}$ devices we use MoS$_2$ monolayers with smooth straight edges. Here the twist angle is defined as the relative orientation between the straight edges of top and bottom flakes by using optical microscopy. With this definition, we fabricated twisted devices mostly near $\theta\to0^\circ$ and a few near $\theta\to60^\circ$. Electrical contacts were designed by e-beam lithography, followed by 5/50nm Cr/Au deposition. Room temperature gate voltage dependence of Raman measurements were performed with 532 nm laser excitation under high vaccum ($10^{-5}$mbar). Laser power was kept below 1.5 mW to avoid sample heating. For gating, we have used 285 nm SiO$_2$ back gate. The threshold voltage ($V_{Th}$), at which the device switches from the off to the on state was determined (see Supporting information, Fig.~S3). The effective electron doping concentration was determined using the equation, $ne=C_{ox}(V_{g}-V_{Th})=C_{ox}\tilde{V}_{g}$ ,where $C_{ox}$ is the gate capacitance per unit area (here 1.2$\times10^{-4}$ F/m${^2}$). The maximum electron doping that we have studied in this work is $\sim 0.004 e/\mathrm{cell}$ at $\tilde{V}_{g}=60$ V. 

\begin{figure*}[t]
	\vspace*{-0.3cm}
	\hspace*{-0.5cm}
	\includegraphics[width=1.05\linewidth]{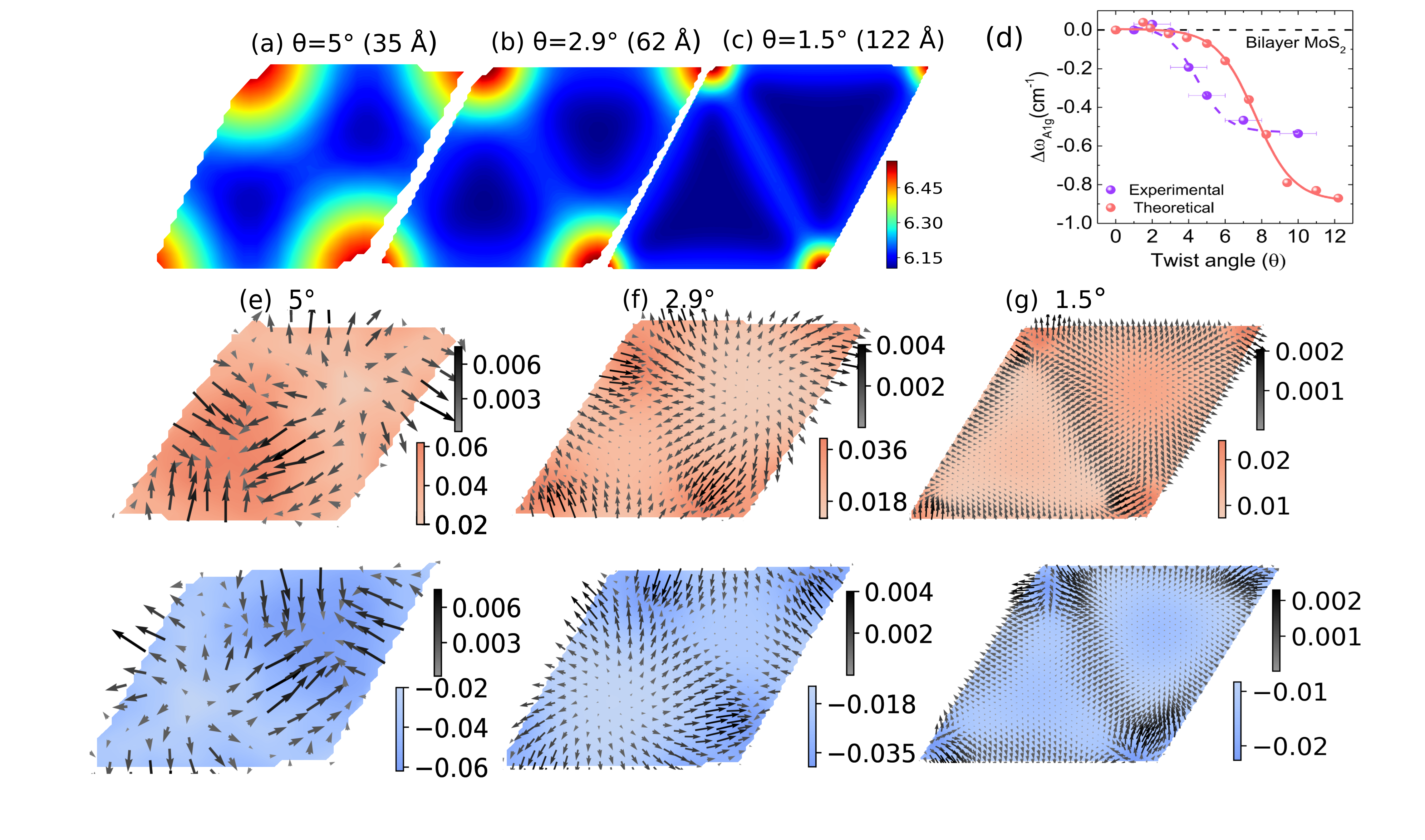}
	\caption{(a)-(c): Evolution of ILS landscape of the relaxed $\mathrm{tBLMoS_{2}}$ with moir{\'e} lattice constants written in brackets. The colorbar denotes ILS in {\AA}. (d): Twist angle dependence of $\mathrm{\Delta \omega_{A_{1g}}}$ mode. Both the theoretical and experimental data are fitted with sigmoid function (see main text for fitting parameters). (e),(f),(g): A visualization of the vibrations of bottom S atoms of top layer $\mathrm{MoS_{2}}$ in the $\mathrm{tBLMoS_{2}}$ correspoding to the normalized $\mathrm{A_{1g}}$ mode for $\theta=5^{\circ},2.9^{\circ}, 1.5^{\circ}$, respectively (Top panel). Similarly, the panel below shows the corresponding eigenvectors for top S atoms of bottom layer $\mathrm{MoS_{2}}$. The out-of-plane displacements are denoted as colored field, whereas arrows denote in-plane displacements.}
	\label{Fig3}	
\end{figure*}

To investigate the twist angle dependence of the interlayer interaction in $\mathrm{tBLMoS_{2}}$ we measured the Raman spectra of $\mathrm{tBLMoS_{2}}$ for several twist angles (Fig.~1c) with focus on two prominent first-order vibrational modes,  in-plane $\mathrm{E^1_{2g}}$ and  out-of-plane $\mathrm{A_{1g}}$ mode. It should be noted that, the $\mathrm{E^1_{2g}}$ redshifts and $\mathrm{A_{1g}}$ blueshifts as we increase the layer number from mono layer to bilayer $\mathrm{MoS_{2}}$ \cite{lee2010anomalous,molina2011phonons}. As is well known, the blueshift of the $\mathrm{A_{1g}}$ mode with increasing layer number is due to additional ``springs" between two neighboring $\mathrm{MoS_{2}}$ layers, whereas the redshift of the $\mathrm{E^{1}_{2g}}$ mode frequencies arises from enhanced dielectric screening of Coulomb interaction \cite{molina2011phonons}. The high-symmetry stacking regions present in the $\mathrm{tBLMoS_{2}}$ are different for $\theta\to0^\circ$, and $\theta \to60^\circ$. This is due to the presence of different sub-lattice atoms (Mo, S) in the unit-cell \cite{Maity_ultrasoft_2019,naik2018ultraflatbands}. Here, we mainly focus on twist angles near $0^\circ$. Among all the possible stackings $\mathrm{AA^{\prime}}$ stacking ($\theta=60^{\circ}$, Mo, S of top layer are directly above S, Mo of bottom layer) is known to be most stable. As a result, naturally exfoliated $\mathrm{BLMoS_{2}}$ has $\mathrm{AA^{\prime}}$ stacking. In Fig.~1d we show the evolution of $\mathrm{E^1_{2g}}$, and $\mathrm{A_{1g}}$ mode frequencies with several twist angles. It is clear that as $\theta$ decreases (from $10^{\circ} \to 1^{\circ}$) the $\mathrm{E^1_{2g}}$ mode redshifts while $\mathrm{A_{1g}}$ blueshifts \textit{monotonically}. The separation between these peaks, ($\omega_{\mathrm{A_{1g}}}-\omega_{\mathrm{E^{1}_{2g}}}$) are often used as a quantitative measure of the strength of the interlayer mechanical coupling \cite{liu2014evolution}. For brevity, we refer interlayer mechanical coupling as interlayer coupling. The larger the peak separation stronger is the interlayer coupling. For example, we find the peak separation of the $\mathrm{BLMoS_{2}}$ ($\sim 21.3 \ \mathrm{cm^{-1}}$) is always larger than that of $\mathrm{SLMoS_{2}}$ ($\sim 18.7 \ \mathrm{cm^{-1}}$). Remarkably, we discover that the peak separation can be tuned with twist angles in the $\mathrm{tBLMoS_{2}}$ \textit{monotonically} (Fig~1e). The peak separation is maximum for  $\theta \lesssim 2^\circ$ (saturates to  $\mathrm{BLMoS_{2}}$ value) and minimum for $\theta \sim 10^{\circ}$ (slightly greater than $\mathrm{SLMoS_{2}}$ value). This immediately implies the tunability of the interlayer coupling strength with twist angles. Moreover, it also suggests that $\mathrm{tBLMoS_{2}}$ with small (large) twist angle behaves like $\mathrm{BLMoS_{2}}$ ($\mathrm{SLMoS_{2}}$).

Next, we investigate the evolution of the high-frequency phonon modes in $\mathrm{tBLMoS_{2}}$ upon electron doping. A schematic and micrograph of $\mathrm{tBLMoS_{2}}$ devices of this experiment are shown in Fig.~2a and 2b, respectively. In Fig.~2c we show the gate voltage dependence of $\mathrm{A_{1g}}$ and $\mathrm{E^1_{2g}}$ modes for 7$^{\circ}$ twist angle. It is evident from the figure that, as gate voltage increases, the $\mathrm{A_{1g}}$ mode gets softened and broadened, whereas the $\mathrm{E^1_{2g}}$ mode remains unaffected. In Fig.~2d,~2e (Fig.~2g,~2h) we show the shift of the $\mathrm{A_{1g}}$ ($\mathrm{E^1_{2g}}$) mode frequency and the corresponding linewidth respectively, as a function of electron doping at different twist angles. Three lorentzians are fitted to the data to obtain line shape parameters, where peak position and full width at half maximum (FWHM) denote phonon frequency and linewidth, respectively (see Supporting information, Fig.~S2 ). Here, $\Delta\omega$ and $\Delta \mathrm{FWHM}$ signify the shift of mode frequency and FWHM, with respect to zero doping (for instance, $\Delta\omega$=$\omega({\tilde{V}_g}\neq0)-\omega({\tilde{V}_g}=0)$). We find that the softening of $\mathrm{A_{1g}}$ mode  of $\mathrm{BLMoS_{2}}$ under electron doping is smaller compared to that of $\mathrm{SLMoS_{2}}$. This can be qualitatively understood by noting that, upon electron doping different conduction valleys with different band curvature are occupied in $\mathrm{SLMoS_{2}}$ and $\mathrm{BLMoS_{2}}$ (K for monolayer, $\mathrm{\Lambda}$ for bilayer Fig.~2i, see Supporting Information for details, Fig.~S4). The rate of the softening of the $\mathrm{A_{1g}}$ mode with $\tilde{V}_g$ is shown in Fig.~2f for different twist angles. For large $\theta$ ($\approx 7^{\circ}$), the slope of softening of $\mathrm{A_{1g}}$ mode frequencies ($\sim -0.016\ \mathrm{cm^{-1}/V}$) are identical to that of $\mathrm{SLMoS_{2}}$. This further suggests that, two layers are very weakly coupled for large twist angles. As $\theta \to 0^{\circ}$ ($\theta \lesssim 2^{\circ}$), the softening of the $\mathrm{A_{1g}}$ resembles to that of $\mathrm{BLMoS_{2}}$. In sharp contrast to the $\mathrm{A_{1g}}$ mode, the phonon frequencies of $\mathrm{E^{1}_{2g}}$ remains unaffected due to electron doping. This is due to large electron-phonon coupling corresponding to the $\mathrm{A_{1g}}$ mode as it preserves the symmetry of the $\mathrm{MoS_{2}}$ lattice \cite{chakraborty2012symmetry}. Our results clearly establishes a one-to-one correspondence between the doping concentration and the softening of the $\mathrm{A_{1g}}$ mode frequency of the $\mathrm{tBLMoS_{2}}$. Furthermore, in Fig.~2e and Fig.~2h we show that the change in linewidth ($\Delta$FWHM) of $\mathrm{A_{1g}}$ mode in $\mathrm{tBLMoS_{2}}$ increases significantly under electron doping, whereas the linewidth  $\mathrm{E^{1}_{2g}}$ mode remains unaffected. It should be noted that, the change in $\Delta$FWHM within the applied range of gate voltages are similar for both $\mathrm{SLMoS_{2}}$ and $\mathrm{BLMoS_{2}}$. Interestingly, the change of $\Delta$FWHM for $\theta\approx 1^{\circ}$ is strikingly different (Fig.~2e). We find that, the $\Delta$FWHM not only saturates at smaller gate voltages, but also attains a unmistakably smaller maximum than all other measured twist angles at $\tilde{V}_{g}=60$ V.

 \textit{Simulation:} The introduction of twist in bilayer $\mathrm{MoS_{2}}$ leads to the co-existence of multiple high-symmetry stacking regions in the $\mathrm{tBLMoS_{2}}$. To compare with experimental data, we restrict the theoretical analysis only for $\theta \to 0^{\circ}$. In an unrelaxed $\mathrm{tBLMoS_{2}}$ with $\theta \to 0^{\circ}$, we find two unique high-symmetry stacking regions, AA (where Mo, S of top layer are directly above Mo, S of bottom layer) and AB (Bernal stacking with Mo of top layer directly above S of bottom layer, same as BA with Bernal stacking having S of top layer are directly above Mo of bottom layer) {\cite{liang2017interlayer}. Among these stackings, AA (AB, BA) stacking is the energetically most unfavorable (favorable) \cite{Mit_kc_2019, Maity_ultrasoft_2019}. Upon relaxing these structures, the most stable stacking region grow significantly. In order to illustrate this, we show the evolution of interlayer separation (ILS) landscape in Fig~3a-c, calculated from the separation of Mo atoms of different $\mathrm{MoS_{2}}$ layers \cite{Maity_ultrasoft_2019}. As $\theta$ decreases the stable stacking regions AB, BA (alternate blue triangles with minimum ILS) occupy significantly large area-fraction of the $\mathrm{tBLMoS_{2}}$ than the unfavorable AA stacking region (red circles with maximum ILS). This twist angle dependence of the ILS landscape inherently controls the evolution of high-frequency Raman modes. 

 \begin{figure}[t]
 	\vspace*{-0.5cm}
 	\includegraphics[width=\linewidth]{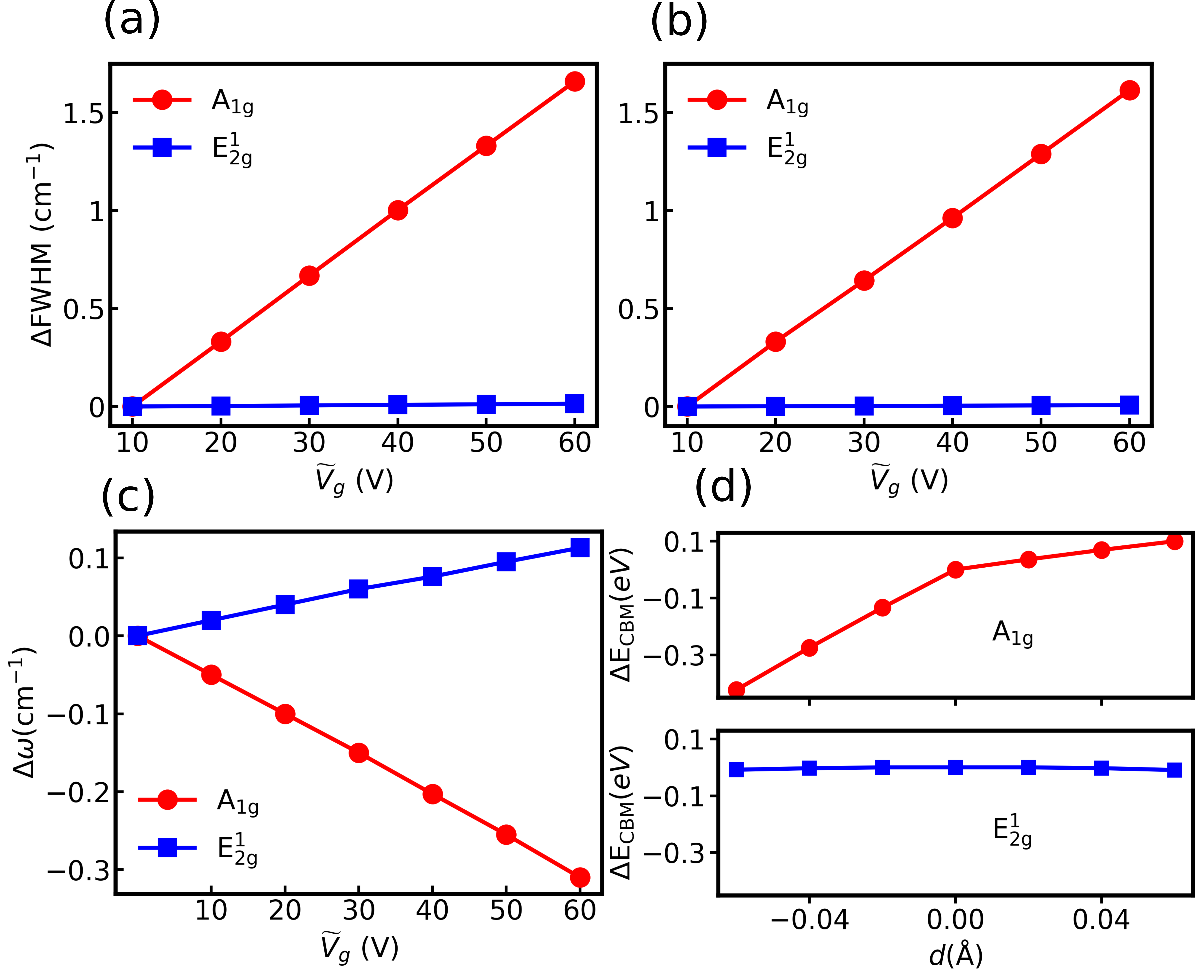}
 	\caption{(a),(b) The doping dependence of linewidth of the $\mathrm{E^{1}_{2g}}$, $\mathrm{A_{1g}}$ mode for single layer and bilayer $\mathrm{MoS_{2}}$, respectively. (c) The doping dependence of the phonon mode frequencies for single layer $\mathrm{MoS_{2}}$. (d) The change of the conduction band minimum at K point of electronic band structure due to different phonon modes with several displacement amplitude (large amplitude is used to show the difference clearly).}
 	\label{Fig3}	
 \end{figure}

Considering the significantly large cost of phonon calculations using first principles based methods, we adopt classical forcefield based simulations to compute the $\mathrm{A_{1g}}$ mode frequencies for $\mathrm{tBLMoS_{2}}$. In Fig~3d we show the evolution of the $\mathrm{A_{1g}}$ mode frequencies (with respect to bilayer $\mathrm{A_{1g}}$ mode) for the relevant twist angles of our experiment. The calculated $\mathrm{A_{1g}}$ mode is only shown with the largest projection, $\mathrm{p^{A_{1g}}}$. The projection is defined as, $\langle \hat{e}_\mathrm{tBLMoS_{2}}|\hat{e}_{\mathrm{A_{1g}}}\rangle$, where the eigenvectors of the $\mathrm{tBLMoS_{2}}$ are projected on the $\mathrm{A_{1g}}$ mode of $\mathrm{BLMoS_{2}}$. Our calculations correctly capture the \textit{monotonic} increment of the $\mathrm{A_{1g}}$ mode frequencies as $\theta \to 0^{\circ}$. Furthermore, we find both the theoretical and experimental data can be fitted well with a sigmoid curve, $A/(1+e^{(-(\theta-\theta_{0})/d))}$ for $\Delta \omega_{\mathrm{A_{1g}}}$ (Fig.~3d). The fitted parameters are also consistent with each other (theoretical : $A=-0.88 \pm 0.02$, $\theta_0=7.6\pm0.1$, $d=1\pm0.1$; experimental : $A=-0.55 \pm 0.02$, $\theta_0=4.5\pm0.2$, $d=0.8\pm0.2$). It should also be noted that, the experimentally assigned twist angle has an uncertainty of $1^{\circ}$ (Fig.~3d). In order to understand the twist angle dependence of the $\mathrm{A_{1g}}$ mode from microscopic view point, we show the eigenvector components on the interlayer-nearest-neighbor S atoms in the $\mathrm{tBLMoS_{2}}$ (Fig~3e,3f,3g). For large $\theta$, the neighboring S atoms from two layers \textit{locally} move out-of-phase but not by same amplitude (``incoherent", Fig~3e). This is due to the presence of multiple high-symmetry stacking regions of equal area-fraction (Fig~3a), leading to inhomogeneous interlayer coupling \cite{Maity_ultrasoft_2019}. As $\theta \to 0^{\circ}$, the neighboring S atoms from two layers \textit{locally} move out-of-phase and by similar amplitude (``coherent", Fig~3g). This is due to the significant increment of the stable AB, BA stacking area as shown in the ILS landscape (Fig~3c). The amplitudes of the in-plane displacement are always one order of magnitude smaller than the out-of-plane displacements. Both the ILS landscape and evolution of the $\mathrm{A_{1g}}$ mode clearly justify the bilayer like behavior for $\mathrm{tBLMoS_{2}}$ with small twist ($\theta \lesssim 3^{\circ}$) and single layer like behavior at large twist angles observed in our experiment.

A similar detailed quantitative estimate for the twist angle dependence of the $\mathrm{E^{1}_{2g}}$ mode frequencies require proper treatment of the dielectric screening of Coulomb interaction \cite{molina2011phonons}. The present parametrization of the classical forcefield used in our calculation does not take into account this screening properly. Hence, we qualitatively explain the observed trend for the $\mathrm{E^{1}_{2g}}$ mode using first principles calculations based on density functional perturbation theory (DFPT). Our argument relies on two important observations of the $\theta$ dependence of the relaxed ILS landscape : (i) For $\theta \lesssim 3^{\circ}$, AB/BA occupy significantly large area-fraction of the $\mathrm{tBLMoS_{2}}$ and the ILS at these stackings are minimum ($\approx$6.1 \AA), (ii) For relatively large twist angles both the stable stackings not only occuply similar area-fraction as that of AA but also are at a larger ILS ($\gtrsim 6.2$ \AA)\cite{Maity_ultrasoft_2019}. Thus, by simply noting the change in $\mathrm{E^{1}_{2g}}$ mode frequencies of AB stacking by increasing the ILS from their minimum, we can qualitatively understand the $\theta$ dependence. We note that, the ILS at AB decreases as $\theta \to 0^{\circ}$ and saturates to it's minimum for $\theta \lesssim 3^{\circ}$ \cite{Maity_ultrasoft_2019}. We increase the ILS of AB by $\sim 0.3$ {\AA} from their equilibrium spacing, which causes stiffening of the $\mathrm{E^{1}_{2g}}$ mode by $\sim 0.2$ $\mathrm{cm^{-1}}$. In sharp contrast, the $\mathrm{A_{1g}}$ mode softens by $\sim 3 \ \mathrm{cm^{-1}}$. Frequencies of both the modes become closer to their single layer values with increment in ILS. In $\mathrm{SLMoS_{2}}$, the frequency of the $\mathrm{E^1_{2g}}$ ($\mathrm{A_{1g}}$) mode is greater (lesser) than that of $\mathrm{BLMoS_{2}}$ ~ \cite{molina2011phonons, lee2010anomalous}. Thus, the frequency shifts in the $\mathrm{tBLMoS_{2}}$ can be qualitatively understood in terms of the change of the ILS.

 The intrinsic linewidth of the phonon modes can have two origins, phonon-phonon anharmonic effects and EPC. In view of our doping-dependent Raman measurements, we study only the linewidth change due to EPC. In our calculations of linewidth, the electron doping is modelled in the rigid band approximation by shifting the Fermi level above the conduction band minimum \cite{Caruso_prl_2017}. In Fig~4a,4b we show the calculated change in linewidth (FWHM) of both $\mathrm{SLMoS_{2}}$ and $\mathrm{BLMoS_{2}}$ for the relevant doping concentration in our experiment. The FWHM of the $\mathrm{A_{1g}}$ mode increases significantly more than that of the $\mathrm{E^{1}_{2g}}$ mode for both $\mathrm{SLMoS_{2}}$ and $\mathrm{BLMoS_{2}}$. This clearly indicates strong EPC for $\mathrm{A_{1g}}$ mode, which is in excellent agreement with our experiment. The ratio of the EPC strength at $\Gamma$ also confirms this conclusion ($\lambda_{\mathrm{A_{1g}}} / \lambda_{\mathrm{E^{1}_{2g}}} \approx 10^{2}$). The computed change of FWHM ($\sim 1.3$ cm$^{-1}$ with GGA) are also in good agreement with our experiment ($\sim 1.2 \ \mathrm{cm^{-1}}$). Since the change of FWHM due to doping can depend on the exchange-correlation functional used, we confirm our conclusions using LDA as well (see SI).

 We also compute the change in the phonon mode frequencies due to electron doping within the Born-Oppenheimer approximation using DFPT. The electron doping corresponding to different gate voltages is simulated by adding a fraction of electrons in the $\mathrm{SLMoS_{2}}$ unit cell. We find the $\mathrm{A_{1g}}$ mode softens significantly, while the change in $\mathrm{E^{1}_{2g}}$ mode frequency is negligible (Fig~4c). It is well known that, the magnitude of the softening can be sensitive to exchange-correlation used \cite{novko2019broken,Ponomarev_multivalley_2019}. The softening with GGA $\sim 0.3 \ \mathrm{cm^{-1}}$ (LDA $\sim 0.9 \ \mathrm{cm^{-1}}$) is underestimated (overestimated, see SI) than our experimental results $\sim 0.75 \ \mathrm{cm^{-1}}$ for $\tilde{V}_{g}=50 \ V$. For $\mathrm{BLMoS_{2}}$, we find the softening of the $\mathrm{A_{1g}}$ mode frequency ($\sim 0.15\ \mathrm{cm^{-1}}$) is marginally lower than that found for $\mathrm{SLMoS_{2}}$ with GGA, consistent with the trends in our experiment. However, this small change is within the accuracy of our calculations. 

 In order to qualitatively understand the strong (weak) EPC for the $\mathrm{A_{1g}}$ ($\mathrm{E^{1}_{2g}}$) mode, we study the change in the electronic band structure due to frozen atomic displacement corresponding to the phonon mode \cite{bardeen1950deformation,khan1984deformation}. For $\mathrm{SLMoS_{2}}$, from the Fermi level position for relevant electron doping in our experiment we conclude only K valley near conduction band minimum (CBM) is populated. We find CBM at K point changes significantly due to $\mathrm{A_{1g}}$ mode implying a strong EPC, whereas it remains practically unchanged due to $\mathrm{E^{1}_{2g}}$ mode implying weak EPC (Fig~4d). This simple argument can also be easily extended for the $\mathrm{BLMoS_{2}}$, confirming our experimental observations.

The quantitative estimates of the softening of the phonon modes and FWHM with twist angles due to doping is computationally challenging. Nevertheless, the doping dependence of the softening of the $\mathrm{A_{1g}}$ mode (single layer like at large $\theta$, bilayer like at small $\theta$, Fig~2f) can be qualitatively understood from the evolution of interlayer coupling presented earlier. Although, the $\Delta$FWHM for $\theta\approx1^{\circ}$ (Fig~2e) saturates at significantly lower gate voltage unlike $\mathrm{BLMoS_{2}}$, which can not be explained without explicit calculation of EPC and probably hints to band flattening. It is also interesting to note that, at larger electron doping than that considered here, multiple inequivalent valleys might be occupied, which can lead to greater softening of the phonon mode frequencies, superconductivity \cite{Ponomarev_multivalley_2019,novko2019broken,piatti2018multi,ge2013phonon,costanzo2016gate}.

The low-frequency shear and layer breathing modes originate from the relative displacement of the constituent layers in bilayer $\mathrm{MoS_{2}}$ and provides a non-destructive probe to the interlayer coupling \cite{maity2018temperature,zhao2013interlayer, liang2017low}. Therefore, these low-frequency modes can further provide insights into the evolution of the interlayer coupling strength and can also be used as a sensitive probe for twist angle \cite{Maity_ultrasoft_2019,huang2016low,puretzky2016twisted}. Furthermore, recent computational study also suggests the existence of the phason modes in all twisted bilayer structures\cite{Maity_ultrasoft_2019}. The detection of these phason modes and their twist angle dependent velocity can provide valuable information of the rigidity of the moir\'{e} lattices.   

\section{Conclusion}
We have demonstrated systematic evolution of interlayer coupling in twisted bilayer $\mathrm{MoS_{2}}$ using Raman spectroscopy. When the twist between two layers is large, $\mathrm{tBLMoS_{2}}$ behaves like $\mathrm{SLMoS_{2}}$, whereas for small twist angles it behaves like $\mathrm{BLMoS_{2}}$. Furthermore, using doping dependent Raman spectroscopy we discover strong EPC corresponding to the $\mathrm{A_{1g}}$ mode irrespective of the number of layers or twist angle between them, unlike $\mathrm{E^{1}_{2g}}$ mode that shows weak EPC. We explain our results by combining classical forcefield and first principles based simulations. Our study provides another step toward twistnonics \cite{Maity_ultrasoft_2019} and can be generalized to other vdW heterostructures.

\section{Materials and Methods}
\subsection{Experiment}
\textbf{Dry transfer method:} MoS$_2$ flakes were exfoliated on different SiO$_2$/($p^+$)Si substrates. A drop of PDMS polymer was placed on a transparent glass slide and baked at $150^\circ$~C for 30min. We used LCC polymar for sacrificial layer, that spin coated on PDMS drop at 8K rpm and baked for 3 hrs. The glass slide containing the PDMS drop, facing down, was aligned with the substrate containing the flake under the microscope. Using a micromanipulator, the alignment was made in such a way, that, when the glass slide and the substrate are brought closer, the convex surface of the LCC touches the substrate only over the region surrounding the flake. Following a contact time of 5 mins at $57^\circ$~C, the glass slide is retracted, lifting the flake from the substrate due to the strong adhesion to LCC. To avoid the deformation of LCC and tearing the flakes, the lower substrate temperature was decreased to $50^\circ$~C before detaching from LCC. In the next step, the LCC drop, containing the upper layer flake, is precisely aligned manually by using micromanipulator at desired twist angle with the bottom flake, and picked up in the similar manner to form the heterostructure. In the final step, the glass slide containing the LCC, along with the heterostructure, is pressed against a RCA cleaned patterned SiO$_2$/($p^+$)Si at $120^\circ$~C. The desired heterostructure sticks to the patterned SiO$_2$/($p^+$)Si substrate along with part of LCC. Later, LCC is dissolved in acetone to get the heterostructures on patterned SiO$_2$/($p^+$)Si substrate.

\textbf{Raman spectroscopy:} Raman spectra of MoS$_2$ flakes were done by using Horiba LabRAM HR spectrometer. 532nm laser with laser power less than 1.5mW was used for spectroscopy measurement. For gate voltage dependence Raman spectroscopy vacuum compatible($10^{-5}$mbar) chamber was used. 

\subsection{Theoretical Calculations}
\textbf{Classical Simulations:} We use the Twister code \cite{naik2018ultraflatbands} to create twisted bilayer $\mathrm{MoS_{2}}$ for several commensurate twist angles.   The commensurate twist angles and corresponding moir\'{e} lattice constants are calculated in the following manner : $\theta=\cos^{-1}\big(\frac{m^2 + 4mn + n^2}{2(m^2 + mn + n^2)}\big)$ , and  $L_{\mathrm{m}}=\frac{|m-n|a}{2\sin(\theta/2)}$, with $m,n$ are integers and $a$ is unit-cell $\mathrm{MoS_{2}}$ lattice constant. The smallest moir\'{e} lattice constants are only found for $|m-n|=1$, which are feasible to simulate. We use the Stillinger-Weber and Kolmogorov-Crespi potential to capture the intralayer and interlayer interactions, respectively \cite{Plimpton_jcp_1995, Jiang_iop_2015,Mit_kc_2019}. The phonon frequencies are calculated on the relaxed $\mathrm{tBLMoS_{2}}$ using PHONOPY\cite{Togo_sm_2015}. 

\textbf{Quantum Simulations:} The first principles based quantum calculations using density functional theory \cite{kohn1965self,hohenberg1964inhomogeneous} are performed with both local density approximation (LDA) and  Perdew-Burke-Ernzerhof generalized gradient approximation (GGA) for the exchange-correlation functional as implemented in Quantum ESPRESSO \cite{giannozzi2009quantum}. We have used a plane wave basis set with a kinetic energy cutoff of 80 Ry, optimized norm conserving Vanderbilt pseudopotential \cite{hamann2013optimized}, and a $12 \times 12 \times 1$ uniform $\vec{k}$ point Monkhorst-Pack grid for the sampling of the Brillouin zone with 30 {\AA} vacuum. For the GGA we use GGA+DFT-D \cite{grimme2006semiempirical} for the $\mathrm{BLMoS_{2}}$ calculations. For the linewidth calculations, we compute the phonon frequencies and eigenvectors using a $6 \times 6 \times 1$ uniform $\vec{q}$ point grid and determine maximally localized wannier functions\cite{marzari2012maximally,mostofi2008wannier90} with much finer electron grid ($72\times72\times1 \ \vec{k}$), phonon grid ($72\times72\times1 \ \vec{q}$) to interpolate electron-phonon coupling matrix elements with EPW \cite{giustino2007electron,ponce2016epw}. The lineiwdth is finally calculated from the imaginary part of the phonon self energy using EPW with $10^{6}$ $\vec{k}$ points. In the double delta approximation, the smearing used to replace the delta function is chosen to be relatively large (0.1 eV). The electron doping for the linewidth calculations is simulated within the rigid band approximation by shifting the Fermi level above conduction band minimum. The Fermi level is determined by adding a fraction of electron to the unit cell and using fine $1000\times 1000\times 1 \ \vec{k}$ grid with a small smearing to the energy conserving delta function $\sim 0.002$ eV with EPW. 

For the doping dependent phonon calculations (at $\Gamma$) we explicitly add a small fraction of electrons corresponding to the gate voltage in our experiment. For $\mathrm{SLMoS_{2}}$ we use a dense $96\times96\times1 \ \vec{k}$ grid and a smearing of $0.002$ Ry corresponding to room temperature. To correct the boundary conditions for 2D materials we have also computed the phonon mode frequencies by truncating the Coulomb interaction in the out-of-plane direction of $\mathrm{SLMoS_{2}}$. We do not find significant change in the renormalization of phonon mode frequencies. The spin-orbit coupling is not included in our calculations.

\section*{Acknowledgements}

The authors thank the financial support from the Department of Science and Technology, Government of India and Supercomputer Education and Research Center (SERC) at IISc for providing computational resources.

\maketitle
\bibliography{mybib.bib}

\newpage

\section{Supporting Information}
\section{Section 1: Transport characteristics of twisted bilayer MoS$_2$} 

\begin{figure*}[h]
	\includegraphics[width=1\linewidth]{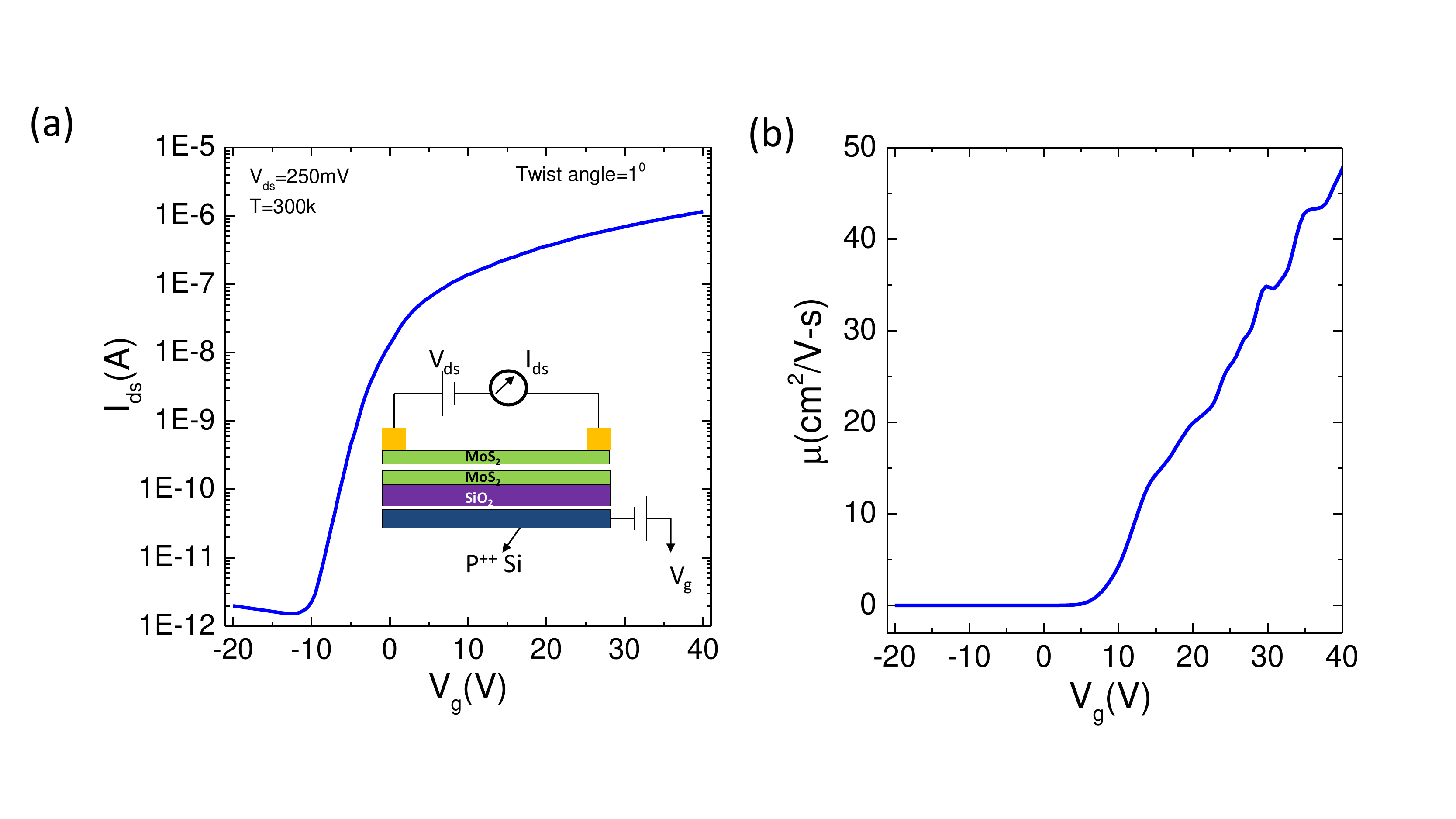}
	\caption{\textbf{Transport characteristics of twisted bilayer MoS$_2$.} (a) $I_{ds}$ (source-drain current) as a function of $V_g$ (back gate voltage) at $V_{ds}$=250mV (source-drain voltage). A typical schematic of the FET is presented in the inset with circuit diagram. Cr/Au(5/50nm) contacts were lithographically defined followed by thermal evaporation to create source and drain electrodes, and 285nm $SiO_2$ was used as the back gate. The on-off ratio of the device was $\sim 10^5$. (b) Room temperature field-effect mobility($\mu=((1/C)\times(d\sigma/dV_g))$) of the FET at $V_{ds}$=250mV, where C is the gate capacitance per unit area, and $\sigma=((L/W)\times(I/V_{ds}))$ is the linear conductivity at low bias. Here L and W are the length and width of the twisted bilayer MoS$_2$ channel.  } 
\end{figure*}
\newpage

\section{Section 2: Lorentzian curve fit to the Raman spectrum}

\begin{figure*}[h]
	\includegraphics[width=1\linewidth]{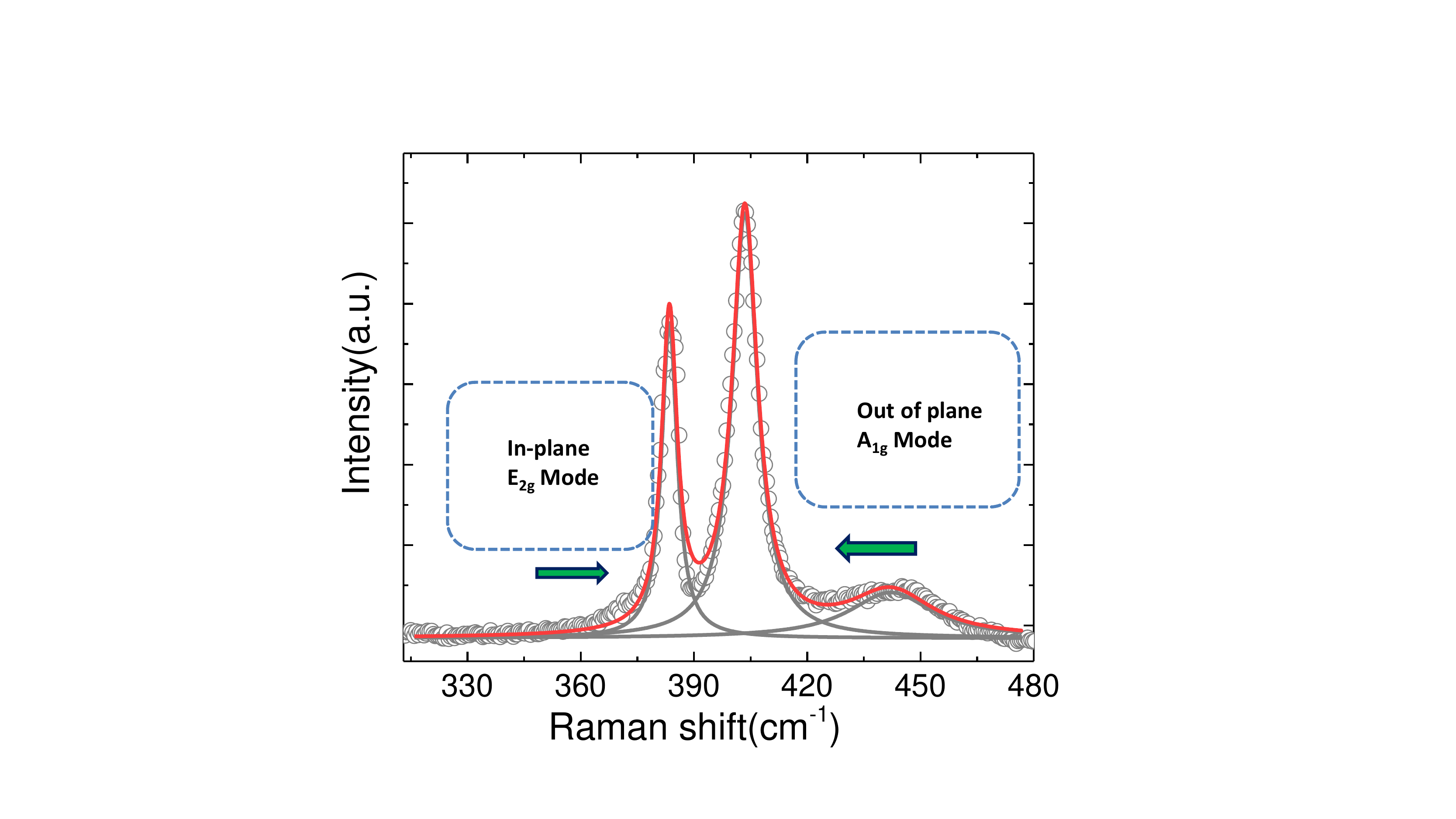}
	\caption{\textbf{Lorentzian curve fit to the Raman spectrum.}   Open circles are the experimental data points, the red line is the Lorentzian fit to the total Raman spectrum, and grey lines are the Lorentzian fit to the individual peak. Line shape parameters (position of the Raman peaks, and corresponding FWHM) are obtained by using sum of three Lorentzians to the data. } 
\end{figure*}
\newpage

\section{Section 3: Determination of the threshold voltage ($V_{Th}$)}

\begin{figure*}[h]
	\includegraphics[width=1\linewidth]{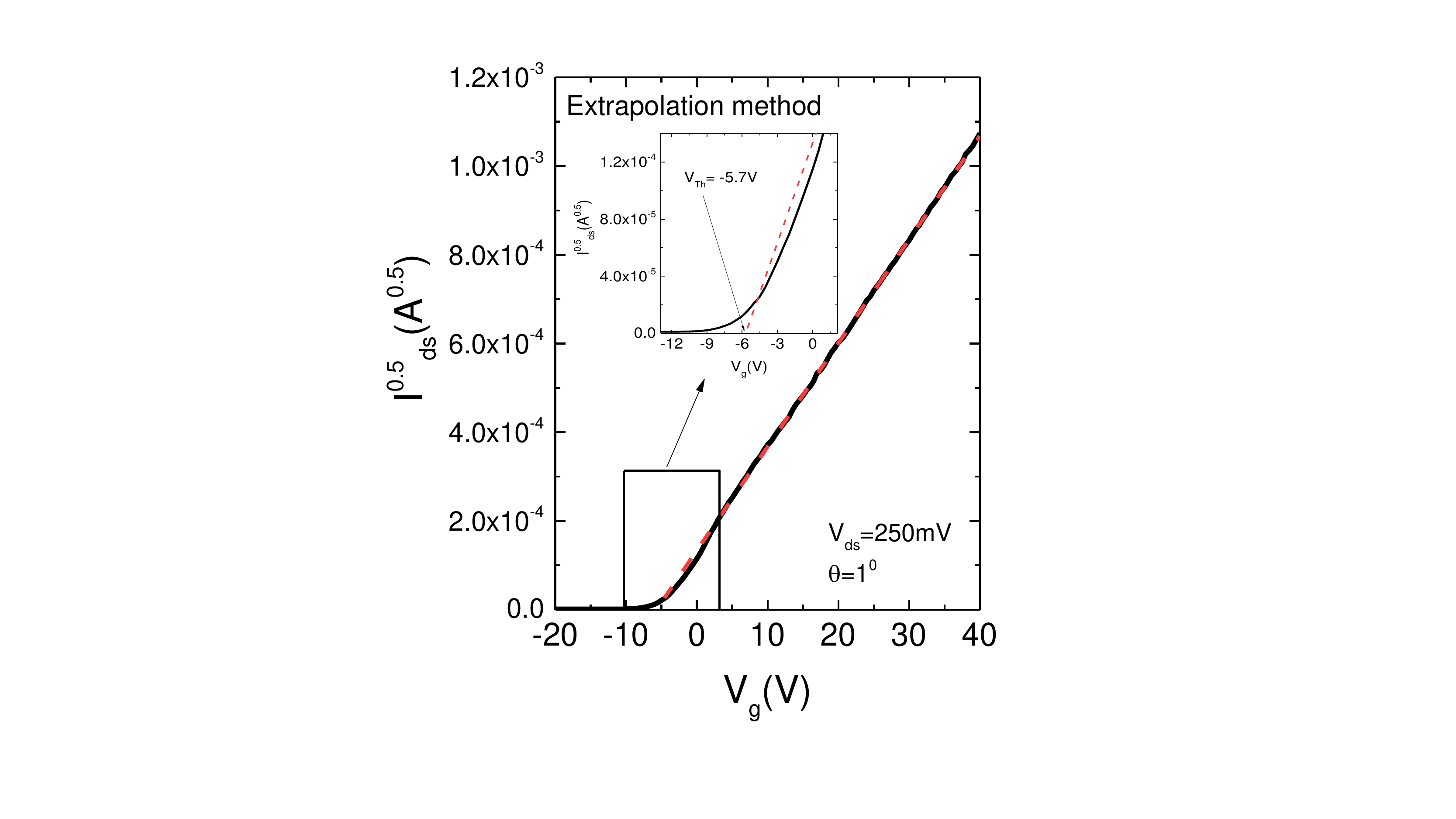}
	\caption{\textbf{Determination of the threshold voltage ($V_{Th}$).} Extrapolation method implemented on the measured $I^{0.5}-V_g$ characteristics of the twisted bilayer MoS$2$ (twist angle, $\Theta=1^\circ$) FET at $V_{ds}$=250mV. This method consist of finding the $V_g$ axis intercept (i.e.,$I^{0.5}$=0) of the linear extrapolation of the $I^{0.5}-V_g$ curve.} 
\end{figure*}
\newpage 

\section{Section 4: Sigmoid curve fitting for $\theta$ dependence of the peak separation}

\begin{figure*}[h]
	\includegraphics[width=1\linewidth]{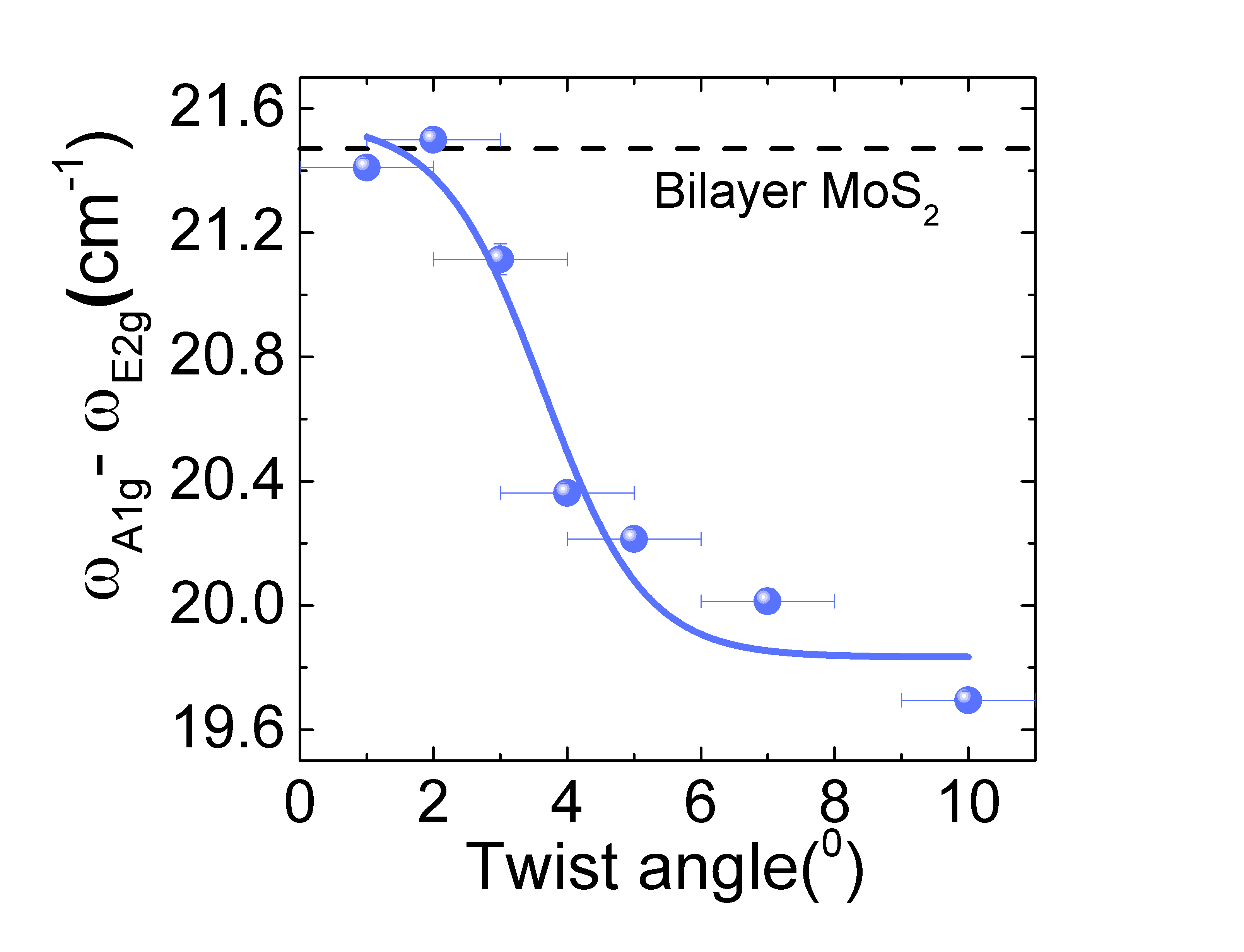}
	\caption{\textbf{Similar to the evolution of $\mathrm{A_{1g}}$ mode frequencies with twist angle, we fit sigmoid function for $\omega_{\mathrm{A_{1g}}}-\omega_{\mathrm{E_{2g}}}$ using the function : $A + (B-A)/(1+e^{(\theta - \theta_{0})/d})$. The fitted parameters $d=0.8\pm0.1$, $\theta_0 = 3.6\pm0.1$ agree well with the parameters for the $\mathrm{A_{1g}}$ mode with $A=19.8, \ B=21.6 $. }} 
\end{figure*}

\newpage 
\section{Section 5: Calculations with local density approximation}
\begin{figure*}[h]
	\includegraphics[width=1\linewidth]{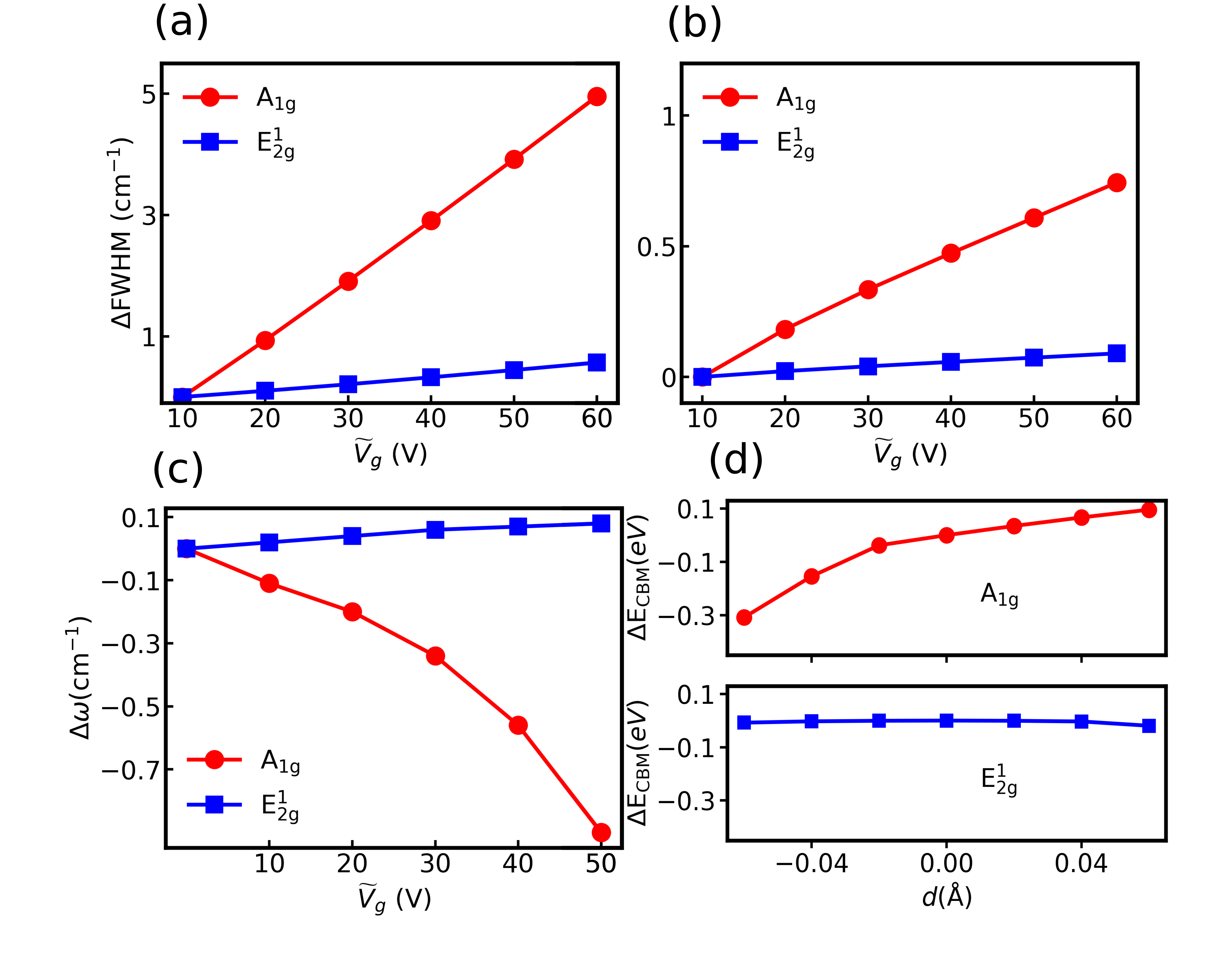}
	\caption{(a),(b) The doping dependence of linewidth of the $\mathrm{E^{1}_{2g}}$, $\mathrm{A_{1g}}$ mode for single layer and bilayer $\mathrm{MoS_{2}}$, respectively. (c) The doping dependence of the phonon mode frequencies for single layer $\mathrm{MoS_{2}}$. (d) The change of the conduction band minimum at K point of electronic band structure due to different phonon modes with several displacement amplitude (large amplitude is used to show the difference clearly).}
\end{figure*} 

Depending on the exchange correlation functional used we find different CBM valleys might be occupied on electron doping for bilayer $\mathrm{MoS_{2}}$. For instance, with LDA (GGA) the CBM in the electronic band structure for $\mathrm{BLMoS_{2}}$ is at $\Lambda$ (K) point. Hence, upon electron doping $\Lambda$ (K) valley will be populated for LDA (GGA). In the $\mathrm{SLMoS_{2}}$, upon electron doping K valley will be populated for both the exchange-correlation functional. This leads to the quantitative difference in linewidth between single and bilayer $\mathrm{MoS_{2}}$ with LDA (whereas almost identical behaviour with GGA). In the main text, the schematic used for different valley occupation in monolayer and bilayer $\mathrm{MoS_{2}}$ was calculated using LDA. More accurate calculation using GW method to compute the band structures suggest that, the trends of different valley occupation captured with LDA is correct. However, the energy separation between the two valleys at $\Lambda,$ K points can only be accurately captured using GW method for both mono layer and bilayer $\mathrm{MoS_{2}}$ (which we leave for future study). The quantitative difference of the doping dependence of the phonon modes found in our experiment arise from the aforementioned valley occupation. In order to illustrate this, we compute the ratio of mono layer to bilayer electronic band curvature near the CBM for both LDA and GGA. We find the ratio with LDA to be $\approx 1.3$, whereas with GGA $\approx 1$. This explains the difference in linewidth found in our calculation with LDA for mono layer and bilayer $\mathrm{MoS_{2}}$.

\end{document}